\documentclass[12pt]{article} 
\title{\bf Persistent current and Drude weight in one-dimensional rings with
substitution potentials}

\author{Giancarlo Queiroz Pellegrino\footnote{e-mail address: gian@dm.ufscar.br}\\
{\normalsize \em Departamento de Matem\'atica, Universidade Federal de S\~ao
Carlos}\\  {\normalsize \em 13560-970 S\~ao Carlos -- SP, Brasil} }

\begin{document} 
\maketitle 
\begin{abstract}

Persistent currents and Drude weights are investigated for the tight-binding
approximation to one-dimensional rings threaded by a magnetic flux and with
potential given by some almost-periodic substitution sequences with different
degrees of randomness, and for various potential strengths. The Drude weight
$D$ distinguishes correctly conductors and insulators, in accordance with
the results shown by the currents. In the case of insulators the decay of
$D(N)$ for large ring lengths $N$ provides an estimate for the localization
length of the system. It is shown that the more random the sequence does
not imply the smaller conducting properties. This discrepancy between the
hierarchy of disorder of the sequences and the capacity of conduction of the
system is explained by the gaps in the energy spectra.\\

\end{abstract}  
\newpage 

\section{Introduction}

The ability of quantum mechanical systems in conducting, and their transport
properties in general, have been greatly considered in recent years in
connection with anomalous transport and localization phenomena \cite{TrPr}.
This consideration has been also intense in the mathematical-physics
community where one would like to know the spectral and state properties of
Schr\"odinger operators -- even one-dimensional, as will always be the case here
-- with potentials lying between the hallmarks set by periodic potentials with
absolutely continuous spectra and Bloch states at one extreme, and random
potentials with pure point spectra and localized states at the other
\cite{MathPhys}.  In between there appears a variety of behaviours which seem to
be dictated in great measure by the randomness of the potential.  To cite just
two properties, it is accepted that the mean square displacement $d(t)$ and the
quantum return probability $C(t)$ for an initial delta-function state are
algebraic functions of time ruled by some exponents, $d(t)\sim t^{2\beta}$
and $C(t)\sim t^{-\alpha }$.  These exponents by themselves reflect the more
or less random the potential is:  ballistic motion in periodic potentials is
characterized by $\beta =\alpha =1$, localization in random potentials by
$\beta =\alpha =0$, while anomalous diffusion in non-periodic potentials shows
$0<\alpha ,\beta <1$. On the mathematical side, a considerable amount of the efforts
have sought to relate these dynamical exponents to the dimensions of the spectral
measure \cite{MathPhys}. However, up to the moment these relations have been very
difficult to be applied in practical calculations: on the one hand, the information
dimension $D_1$ provides only lower bounds for the exponent $\beta$ ruling the
algebraic decay of the mean square displacement; on the other hand, in the case of
the return probability $C(t)$ the associated correlation dimension $D_2$ is not in
general attainable by means of analytical calculations, and its numerical
computation is extremely difficult due mainly to the limiting process involved. In
this scenario one-dimensional non-periodic potentials generated as substitution
sequences have emerged as a convenient theoretical laboratory since the sequences
are neither completely random nor periodic, even though they are formed from well
defined rules.  In addition, they can be classified by their autocorrelation
measures, revealing a hierarchy with respect to disorder.  These two aspects of the
substitution sequences, a sort of correlated disorder, allow us to investigate
transport properties, as well as mathematical ones,  in a more controlled way. In
this spirit it was shown recently the counter-intuitive result that a more random
potential can give rise to better transport properties than a less random one
\cite{OP99}.  From another point of view, with the recent techniques of growing
super-lattices one could think of the experimental and technological exploitation
of systems characterized by these non-periodic substitution potentials \cite{EXP}.

It seems therefore desirable to have results for concrete physical situations, in
order to test theoretical predictions concerning not only disordered systems and
localization phenomena but also the almost-periodic substitution potentials
themselves.  Towards this end, we investigate here the tight-binding approximation
to the problem of mesoscopic rings threaded by a time-independent magnetic flux, and
calculate the persistent current and Drude weight in cases where the on-site
potential is given by some substitution sequences with different degrees of
randomness.  The system is described by the Hamiltonian for a one-dimensional ring
with spinless fermions
\begin{equation}  H=-\sum\limits_{k=1}^N {\left[ {\exp
\left( {i{{2\pi } \over N}\phi } \right)c_k^+c_{k+1}+\exp \left( {-i{{2\pi } \over
N}\phi } \right)c_{k+1}^+c_k } \right]}+\sum\limits_{k=1}^N W_kc_k^+c_k . 
\label{H1} \end{equation} 
In this expression $N$ is the number of sites in a ring with
lattice spacing equal to one, the magnetic flux $\phi$ is measured in units of
the quantum of flux $\phi_0=hc/e$, $W_k$ is the potential energy at site $k$,
and the hopping constant is set equal to one.  This model is particularly
convenient for our purposes for two reasons:  first, it admits non-periodic
substitution sequences in a simple way; second, the association of the wave
vector in the one-dimensional lattice to the parameter $\phi$ allows us to
easily explore the transport properties as functions of the sensitivity of the
energy bands to variations in the magnetic flux.  In this way the Drude weight,
shown by Kohn to be useful as a quantitative characterization of the insulating
state \cite{KOHN64}, and also the persistent current, predicted by B\"uttiker,
Imry and Landauer \cite{BIL83}, can both be calculated as derivatives of the
ground-state energy with respect to the flux.

 It is evident from the model that we assume non-interacting electrons and
zero temperature.  Although electron-electron interactions and average effects
give important contributions to the transport properties --- and to the
amplitude of persistent currents, as we briefly recall below and in section 3
--- these approximations describe quite successfully the qualitative features
of the model \cite{NonInter}.  They are justifiable in the present case since
we attempt mainly to characterize how the randomness of the potential affects
the motion of the particles; in so doing we would like therefore to avoid
the contributions from other effects.

 The aim of this paper is to report and discuss the results, obtained for
Hamiltonian (\ref{H1}), of the persistent current and Drude weight as functions
of the degree of randomness of the potential and for various potential
strengths. Many related works have been published concerning different
aspects of the problem.  Persistent currents, Drude weight, and also optical
conductivity, in mesoscopic rings have been investigated predominantly in
Hubbard models, where the on-site interaction between particles with spin plays a
major role.  This was done in a series of papers ending up in criteria for
determining whether a state is metallic, insulating or super-conducting
\cite{OptCond}.  Interaction between spinless particles together with on-site
potential have been considered, mostly with disordered potentials
\cite{HubDis,BPM94,GS95,SCH98} but also with the Aubry-Andr\'e potential \cite{CH97}.
Recently, particles with spin interacting over finite open chains modulated along
the almost-periodic Fibonacci potential have been considered as well \cite{HI01}. The
general picture emerging from these works is not yet completely clear and points to
a competition between the effects caused by interaction, potential, charge and spin
degrees of freedom; it has been shown, moreover, that the results are sensitive to
whether one averages over realizations of the potential \cite{BPM94,GS95} or whether
one considers individual samples \cite{SCH98}. Concerning solely the effects of the
potential, however, the ring with non-periodic substitution potentials has been much
less considered; Fibonacci potential was treated by Jin {\em et al} \cite{JWHJ97}. 
The properties of systems with substitution potentials have been otherwise
intensively studied in open chains, although mainly in the cases of Fibonacci,
Thue-Morse, and Rudin-Shapiro sequences \cite{OpenChains}.

The Fibonacci potential is by far the best studied one of the cited substitution
potentials. It is generally considered to be the most ordered one either in the sense
of its autocorrelation measure --- pure point as for periodic potentials --- or in
the sense of its dynamical exponents \cite{OP99}. In this paper, however, Fibonacci
potentials will be absent since we would like to compare the results obtained for the
different potentials in half- and quarter-filled rings; recall that the lengths of
the potential sequences in that case are given by the Fibonacci numbers.

 In what follows, section 2 presents the definitions and some properties of the
potentials given by the sequences Thue-Morse, Rudin-Shapiro,
paper-folding and period-doubling.  In section 3.1 persistent currents are
calculated for the various potentials (with different amplitudes) as functions 
of the magnetic flux.  A more quantitative characterization of the
conducting property for the various systems is presented in section 3.2, where
the Drude weight is calculated and where some attempt is made to fit its
behaviour as a function of the length of the ring. Also, in section 3.3 the cases
pointed by the Drude weight to be insulators are analyzed in terms of the gaps in the
energy spectra. Section 4 finishes the article with a discussion of the results
obtained.

\section{Non-periodic substitution potentials}

In the next sections Hamiltonian (\ref{H1}) will be considered with on-site
potential energies $W_k$ given by some sequences which in turn are constructed
using well defined substitution rules.  We will be interested mainly in
almost-periodic sequences which are convenient in the context of disordered
systems because we have not only non-periodicity but also non-perfect
correlation \cite{AG95}.

The sequences we will use are constructed using an alphabet of two letters
$\{a,b\}$ and a specific substitution rule for each sequence:
$$\begin{array}{lll} 
{a\to ab}&{b\to ba}& \ \ \ {\rm Thue-Morse \ (TM)}\cr 
{a\to ab}&{b\to aa}& \ \ \ {\rm period-doubling \ (PD)}.\cr 
\end{array}$$ 
Starting with one of the letters
and applying successively the substitution rules we generate almost-periodic
sequences as, for example, the sequence PD $$abaaabababaaabaa\dots$$ The cases
Rudin-Shapiro (RS) and paper-folding (PF) are worked out of an alphabet of four
letters $\{a',b',c',d'\}$, with the rules 
$$\begin{array}{lllll} 
{a'\to a'b'} & {b'\to a'c'} & {c'\to d'b'} & {d'\to d'c'} & \ \ \ {\rm Rudin-Shapiro \ (RS)}\cr
{a'\to a'b'} & {b'\to c'b'} & {c'\to a'd'} & {d'\to c'd'} & \ \ \ {\rm paper-folding
\ (PF)}, 
\end{array}$$ 
and the identifications $a',b'\to a$ and
$c',d'\to b$ in both cases.  The first elements of the sequence RS are
$$aaabaabaaaabbb\dots$$ At each step these sequences have length $N=2^m$ at the
$m$-th iteration. We then define the potential on site $k$ taking $W_k = 0$ if the
$k$-th letter in the sequence is $a$, and $W_k = \lambda$ in case it is $b$; in such
a way the strength $\lambda$ of a given potential can also be varied.

These non-periodic sequences have a classification with respect to their degree
of randomness, which is based on their autocorrelation measure \cite{AG95}.  As is
the case for periodic sequences, PF and PD have pure point autocorrelation
measures; while RS has absolutely continuous autocorrelation measure, as
independent random sequences do.  The TM case lies in an intermediate place
since it has singular continuous autocorrelation measure.  One would expect
that the differences in the degree of randomness of the potentials would
produce different spectral types for the energy levels, and in this way
explain the more or less difficult motion of the particles.  In other words,
one would expect that the more random the potential the smaller the transport
properties.  The point is that all the rigorously studied cases produce
singular continuous spectra \cite{SPECTRA} (RS and PF being open problems
\cite{ALL97}); yet their dynamical properties are very different and, as it seems, do
not respect the hierarchy of disorder proposed by the autocorrelation measures
\cite{OP99}.  In order to have a clearer picture of the situation we would like to
map the role of the potential as a function of its degree of randomness in physical
quantities such as persistent currents and conductivity via the Drude weight.

In what follows the $N$ eigenvalues $E$ and corresponding eigenfunctions
$\psi_E(k)$ are obtained for Hamiltonian (\ref{H1}) using exact diagonalization
for ring lengths from $N=16$ up to $N=256$.  A gauge transformation can
eliminate the flux in the Hamiltonian by transforming the usual periodic
boundary conditions into $\psi _E\left({k+N} \right)=e^{i2\pi \phi }\psi
_E\left( k \right)$.  In this way it was shown that the energy bands associated
to the lattice vector $q=-2\pi \phi /N$ are periodic functions of the flux
$\phi$ \cite{BY61}.  The effect of the potential on the transport properties are
thus seen as the energy bands being more or less flattened out.

\section{Results}

\subsection{Persistent currents}

Since its discovery in 1983, persistent currents have become an important
subject, both theoretically and experimentally, which is still open in what the
amplitude of the currents is concerned.  Particularly difficult to treat has
been the role of the various contributions to it:  electron-electron
interactions, disorder of the potential, statistical effects.  Much of these
issues have been reviewed in Ref.  \cite{K91} (see also, Ref.  \cite{BPM94}).
As predicted by B\"uttiker, Imry and Landauer \cite{BIL83}, in the presence of a
non-vanishing magnetic flux $\phi$, even if it is time-independent, each energy
band $E_n(\phi)$ carries a persistent current proportional to its derivative
with respect to the flux.  At zero temperature, the contributions of the
$M$ occupied levels below the Fermi energy sum up to 
\begin{equation}
I\left( \phi \right)=\sum\limits_{n=1}^{M} {{{\partial E_n} \over
{\partial \phi }}}.  
\end{equation}

Figure 1 shows typical curves for the currents for half- and quarter-filled
rings with the various substitution potentials at strength $\lambda = 0.2$. 
One can notice that the order of increasing amplitudes is kept all along the flux
axis.  Whence in order to estimate a sequence of increasing currents with
respect to the potential disorder, we plot in Fig.  2 the mean currents
$\left\langle {I^2} \right\rangle ^{1/2}$, averaged over the flux period, for
the interval of potential strength $0.2 \leq \lambda \leq 1.0$.  In these
figures the unit of current is the maximum amplitude obtained for null
potential.

It is interesting to note that this hierarchy of amplitudes does not follow
what would be expected if one is guided solely by a measure of the potential
disorder.  Based on the results given by the autocorrelation measure, one could
predict the sequence (PD, PF)/TM/RS for the decreasing current amplitudes
carried by the different potentials.  However, we see that the results obtained
for half-filling suggest otherwise PF/RS/TM/PD, while for quarter-filling
one would set TM/RS/(PF,PD).

\subsection{Drude weight}

In order to give a more quantitative description of the observations above, we 
calculate in what follows the Drude weight for the various situations. As 
prescribed originally by Kohn \cite{KOHN64}, this quantity --- useful in the 
characterization of conducting properties as we will see bellow --- can be 
obtained as the second derivative of the ground-state energy level with respect 
to the lattice vector $q=-2\pi \phi /N$, which in our case reduces to
\begin{equation}
D= N \sum\limits_{n=1}^{M} {\left. {{\partial^2 E_n} \over 
{\partial \phi^{2} }} \right|_{\phi=\phi_{min}}}.  \label{DRUDE}
\end{equation}
This expression is equivalent to that obtained using Kubo linear response
formulae \cite{OptCond}. In the first works following Kohn, $D$ was calculated at
$\phi=0$ and it was  realized that its signal depends on the number of
occupied levels $M$  being even or odd. In this way diamagnetic response of
various systems  were discussed in the literature, as well as the less common
paramagnetic  one for $M=4p$ (see the work of 1991 by Fye {\it et al.} in 
Ref. \cite{OptCond}). However, as was done by Bouzerar, Poilblanc  and
Montambaux \cite{BPM94} and clearly explained by Giamarchi and Shastry 
\cite{GS95}, $D$ describes a situation of equilibrium between the charge 
carriers and should be calculated at the value of $\phi$ where the ground-state
energy has a minimum ($\phi_{min}$ in the expression above). Due to the 
characteristic lengths of our sequences, we adopt here the latter point of  view and
take $M=4p$ for either half- or quarter-fillings, with $M=N/2$  and
$M=N/4$ respectively. In this way $\phi_{min}=1/2$ in  all cases and we avoid 
different kinds of response, thus comparing all systems in the same situation.

The behaviour of $D$ as a function of $N$ gives a criterion for distinguishing 
between conductor and insulator \cite{KOHN64}. We restrict the remarks to our 
one-dimensional case. In metallic conductors $D$ tends to a finite non-zero value 
as $N \to \infty$; whereas for insulators $D$ vanishes in that limit. This 
criterion has a more intuitive description if one realizes that $D$ is inversely 
proportional to the effective mass tensor, which in one-dimension is simply
the  second derivative written above. We present in Fig. 3 the results of
$D(N)$  for the sequences considered in section 2. It is seen
that the overall aspect of $D(N)$  and its rate of decaying in the various
situations confirm the conduction ordering  suggested by the persistent
currents shown in Fig. 2, that is PF/RS/TM/PD  for half-filling and
TM/RS/(PF,PD) for quarter-filling.

 It has been proposed in the context of Hubbard models 
\cite{KOHN64,HubDis,BPM94} that, in general, insulators show an exponential 
decay, $D(N) \propto \exp(-N/\xi)$, governed by the localization length $\xi$.
The behaviours here vary from well  defined insulator (as, for example, in the
case PD) to conductor (as  quarter-filling TM shows itself), with transitions
between these states in some  cases. We attempted therefore a more general
fitting \begin{equation}
D(N) = A + BN^{\gamma} e^{-{{N} \over {\xi}}}     \label{FIT}
\end{equation}
where the parameters $A,B,\gamma$ and $\xi$ passed a $\chi^{2}$-test, and $D(N)$ is
measured in units of the value $D_0$ attained for each filling in the largest ring
with null potential. Some justifications for the fitting procedure are in order.
Concerning  the sizes  of the system, we avoided filling the range between $N=16$ and
$N=256$ with more points because this would mean to take incomplete sequences, since
each iteration in their construction have a definite size (a power of $2$). On the other
hand, going to greater sizes does not alter the results. Concerning the number of
fitting parameters, formula (\ref{FIT}) was chosen to take into account all possible
behaviours, and also to check for meaningless fitting results. We think this strategy
was successful since there was no situation in which the fitting curve could show
dependence on all parameters. In fact, whenever it showed an exponential decay, the
free-constant parameter $A$ was absent. The results are detailed in Table 1. We see
that most of the  cases showed no dependence on parameter $A$, except for PF
half-filling for which $D = A + BN^{\gamma}$, with the exponent $\gamma$ varying from
$-0.29$ to $-0.95$. In this situation, therefore, $D$ shows no dependence on $\xi$,
presenting instead a  polynomial decay. On the other hand, the case TM quarter-filling
did not admit the fitting above.

The fitting curves $D(N)$ appear in Fig. 3 as solid lines, together with the numerical
results obtained from  eq. (\ref{DRUDE}) (filled circles). Dashed lines are for the
cases where the fitting expression given by eq. (\ref{FIT}) does not work. We see
excellent agreement in PD and PF cases, and in TM at half-filling; good agreement  in
RS case, and none in TM quarter-filling. The fitting procedure is interesting
at this point  because it provides information --- via the localization length $\xi$,
when it occurs --- concerning  the link between the insulating property and disorder
under the action of the various  substitution potentials. In figure 4, $\xi$ is plotted
as a function of the potential  strength $\lambda$. As expected the localization length
diminishes with the increase  of potential strength, thus meaning  more localized wave
functions. Obviously if  $\xi > N$ for a given length $N$ the corresponding wave
functions will cover the  whole ring and the system can conduct. The finite value of
$\xi$ suggests, however, that if $N$ is sufficiently large the system will show
itself as an insulator. This is the case for TM and RS at half-filling, and also for
RS at quarter-filling, for potential strengths $\lambda < 0.4$.

\subsection{Spectral gaps}

The characteristics seen in the Drude weights for insulators can be explained in 
terms of the widths of the gaps in the energy spectra. In fact, systems with 
singular continuous spectra, as is commonly the case under almost-periodic 
potentials, show spectra which are Cantor sets; therefore gaps appear with different
widths at different energy values. The question of where (large) gaps will appear has
not yet a definite answer, with some clues having been provided by gap-labelling
procedures \cite{GAPLABEL}. For the cases investigated here, gaps open in accordance
with the results shown by the calculations related to the Drude weight. This is
presented in Fig. 5 where the difference between the last filled energy level and
the next one, $\Delta = E_{M+1} - E_M$,  is computed at $\phi=\phi_{min}$ for the
largest ring considered ($N=256$) and  plotted as a function of the potential
strength $\lambda$. We see two distinct  types of gap. The cases  which are clearly
insulators after the criterion put  forth by the exponential decay of the Drude
weight $D$ have gaps increasing with $\lambda$ and much larger than the gaps
shown by those cases which are conductors for the values of $N$ considered. Also, for
the systems which present greater gaps  in Fig. 5, we have checked the linear
relation $\Delta(\lambda) \sim \xi^{-1}(\lambda)$, not shown here (see the article
of  1991 by Fye {\it et al.} in ref. \cite{OptCond}). There is very good agreement 
for TM and PD at half-filling and also for PD and PF at quarter-filling. The 
exception is RS which appears as an insulator but for which the results concerning
the gaps are  commonly intractable. 

This relation between the Drude weight and the gaps in the spectra can be pursued
further. On the one hand, Fye {\it et al.} \cite{OptCond} obtained results pointing
to the independence of the Drude weight of the boundary conditions being periodic
or open. On the other hand, gap-labelling procedures have shown that for open chains
the integrated density of states ($IDS$) for the TM potential has an open gap at
$IDS=1/2$ and a closed gap at $IDS=1/4$; the PD case presents open gaps at both
$IDS=1/2, 1/4$; and for the cases RS and PF there are no conclusive results
\cite{GAPLABEL}. Accepting that these results for open chains are valid for
sufficiently large rings, and noting that the $IDS$ is directly translated into our
filling factor, we see that the results shown in figures 1 -- 4 are consistent with
gap-labelling predictions. In particular, these predictions would explain why the
current is maximal for TM quarter-filling, since there would be a closed gap at
$IDS=1/4$. On the other way, the vanishing currents for the PF case at
quarter-filling could suggest a closed gap at $IDS=1/4$. 

Although this relation clearly needs more investigation, the observations above may
provide a more intuitive understanding of the relation between the transport
properties of the system and its spectral type, as recalled in the Introduction.

\section{Discussion and conclusion}

The results of the previous sections show that the behaviour of $D(N)$ 
distinguishes correctly the conductors and insulators as suggested by the relative 
amplitudes of the persistent currents. Indeed, the exponential decay of $D(N)$ 
in the case of insulators, and the resulting localization length $\xi$, are 
reinforced by the structure of gaps in the energy spectra. The other cases, 
where that decay is slower than exponential, clearly need more investigation to be
modeled by the behaviour of $D(N)$. Chaves and Satija \cite{OptCond} have
recently conjectured the existence of a new phase state between insulator and
conductor for deterministic aperiodic systems, with a polynomial decay of $D$
as a function of the system size $N$. Our results, seen in the figures and
in the fitting output, points to the same direction although suggesting a
richer behaviour, perhaps with mixtures and transitions between the
possibilities above.

 Some care must be exercised in comparing the results presented in this paper
and other results available in the literature dealing with the relations
between transport properties or diffusion exponents and the degree of
randomness of the potential \cite{TrPr,MathPhys,OP99}. The latter ones mostly
deal with a single particle moving in an open chain, with an initial
delta-function state on a given position. In that case all the energy
eigenstates contribute to the quantum state of the system, contrary to the
situation seen here where only the lowest levels, up to the filling, are
present. Whether is applicable here the result of Fye {\it et al.}
\cite{OptCond} that the Drude weight does not depend on the boundary
conditions is to be investigated more systematically.

 We have thus investigated in this work how different degrees of randomness in
the potential affect the persistent currents and the conducting properties
--- via the Drude weight --- of half- and quarter-filled rings threaded by a
magnetic flux. This was done in the tight-binding approximation with the
potential being given by some almost-periodic substitution sequences (Thue-Morse,
Rudin-Shapiro, paper-folding and period-doubling). The results show that it is by no
means obvious what potential allows better conducting properties, if one is guided
solely by a measure of its degree of randomness as, for example, the autocorrelation
measure of the sequence itself. More specifically, that measure predicts the
following hierarchy of disorder (PD, PF)/TM/RS, while the persistent-current
amplitudes obey PF/RS/TM/PD for half-filling and TM/RS/(PF,PD) for quarter-filling.
A more quantitative estimate, given by the decay of the Drude weight as a function
of the ring length, shows indeed that for half-filling TM and PD systems are
insulators and PF ones can conduct, whereas for quarter-filling PD and PF systems
are insulators and TM ones are conductors (RS systems have not so simple a
classification). That discrepancy is, moreover, confirmed and explained by the gaps
opened in the energy spectra as functions of the potential strength for the various
sequences.\\ \ \\

\noindent
{\bf \large Acknowledgement}

It is a pleasure to acknowledge my indebtedness to C\'esar R. de Oliveira for
introducing me to the subjects of substitution sequences and spectral theory,
and for many helpful discussions. I also thank the hospitality of Institut
f\"ur Physik at the Technische Universit\"at Chemnitz (Germany) where part of
this paper was written. This work was partially supported by the Brazilian
agencies FAPESP and CAPES.

\newpage 
\newcounter{figura} 
\noindent 
{\bf Figure Captions} \hspace*{\parindent} \\ 
\begin{list} {Figure
\arabic{figura}.}{\usecounter{figura} 
\setlength{\rightmargin}{\leftmargin}}

\item Persistent currents for the various substitution potentials at strength
$\lambda = 0.2$ and ring length $N = 256$: (a) half-filling; (b) quarter-filling.

\item Mean current amplitudes as a function of the potential strength:  (a)
half-filling; (b) quarter-filling.

\item Drude weight as a function of the number of sites in the ring
(filled circles): (a) half-filling; (b) quarter-filling. Fitting curves are shown as
solid lines.

\item Localization length $\xi$ as a function of the strength of potential
$\lambda$. Results for RS are shown in the inset for clarity. 

\item Gaps in the energy spectra (symbols) at (a) half-filling and (b)
quarter-filling for flux value $\phi = \phi_{min}$

\end{list}
\vspace*{3cm}
\newcounter{tabela} 
\noindent 
{\bf Table Caption} \hspace*{\parindent} \\ 
\begin{list} {Table
\arabic{tabela}.}{\usecounter{tabela} 
\setlength{\rightmargin}{\leftmargin}}

\item Fitting parameters as in eq. (\ref{FIT}) for the various substitution 
potentials. The indices $h$ and $q$ refer to half-filling and quarter-filling, 
respectively. The localization length $\xi$ is given in units of the (nearest 
integer) number of sites. The symbol --- indicates no dependence of the curve 
on the corresponding parameter, whereas the symbol $-\ -$ points to the 
non-applicability of the fitting expression.

\end{list}

\newpage
\noindent 
{\bf Table 1} \\ \ \\ \ \\ \ \\ \ \\
\begin{tabular}{|c|cccc|cccc|}\hline
Potential & $A_h$ & $B_h$ & $\gamma_h$ & $\xi_h$ & $A_q$ & $B_q$ & $\gamma_q$ & $\xi_q$
\\
\hline \hline
 TM 0.2 & --- & 0.96 & 0.01 & 1935 & - - & - - & - - & - - \\
 TM 0.4 & --- & 0.65 & 0.17 & 150 & - - & - - & - - & - - \\
 TM 0.6 & --- & 0.51 & 0.31 & 53 & - - & - - & - - & - - \\
 TM 0.8 & --- & 0.42 & 0.47 & 25 & - - & - - & - - & - - \\
 TM 1.0 & --- & 0.34 & 0.66 & 14 & - - & - - & - - & - - \\ \hline
 RS 0.2 & --- & 0.96 & 0.01 & 1308 & --- & 0.98 & 0.00 & 1205 \\
 RS 0.4 & --- & 0.89 & 0.03 & 427 & --- & 1.02 & -0.03 & 773 \\
 RS 0.6 & --- & 0.51 & 0.23 & 125 & --- & 1.02 & -0.04 & 316 \\
 RS 0.8 & --- & 0.64 & 0.15 & 105 & --- & 0.73 & -0.06 & 125 \\
 RS 1.0 & --- & 0.09 & 1.03 & 20 & --- & 0.88 & -0.04 & 139 \\ \hline
 PD 0.2 & --- & 0.51 & 0.37 & 36 & --- & 0.51 & 0.32 & 55 \\
 PD 0.4 & --- & 0.74 & 0.36 & 17 & --- & 0.19 & 0.82 & 20 \\
 PD 0.6 & --- & 1.07 & 0.13 & 11 & --- & 0.07 & 1.31 & 12 \\
 PD 0.8 & --- & 4.88 & -0.21 & 9  & --- & 0.03 & 1.68 & 9 \\
 PD 1.0 & --- & 6.47 & -0.23 & 7 & --- & 0.01 & 1.97 & 7 \\ \hline
 PF 0.2 & 0.98 & 0.16 & -0.95 & --- & --- & 0.43 & 0.45 & 31 \\
 PF 0.4 & 0.94 & 0.11 & -0.43 & --- & --- & 0.29 & 0.78 & 13 \\
 PF 0.6 & 0.88 & 0.16 & -0.30 & --- & --- & 0.33 & 0.86 & 8 \\
 PF 0.8 & 0.81 & 0.22 & -0.29 & --- & --- & 0.48 & 0.81 & 6 \\
 PF 1.0 & 0.76 & 0.29 & -0.35 & --- & --- & 0.62 & 0.81 & 5 \\
\hline
\end{tabular}
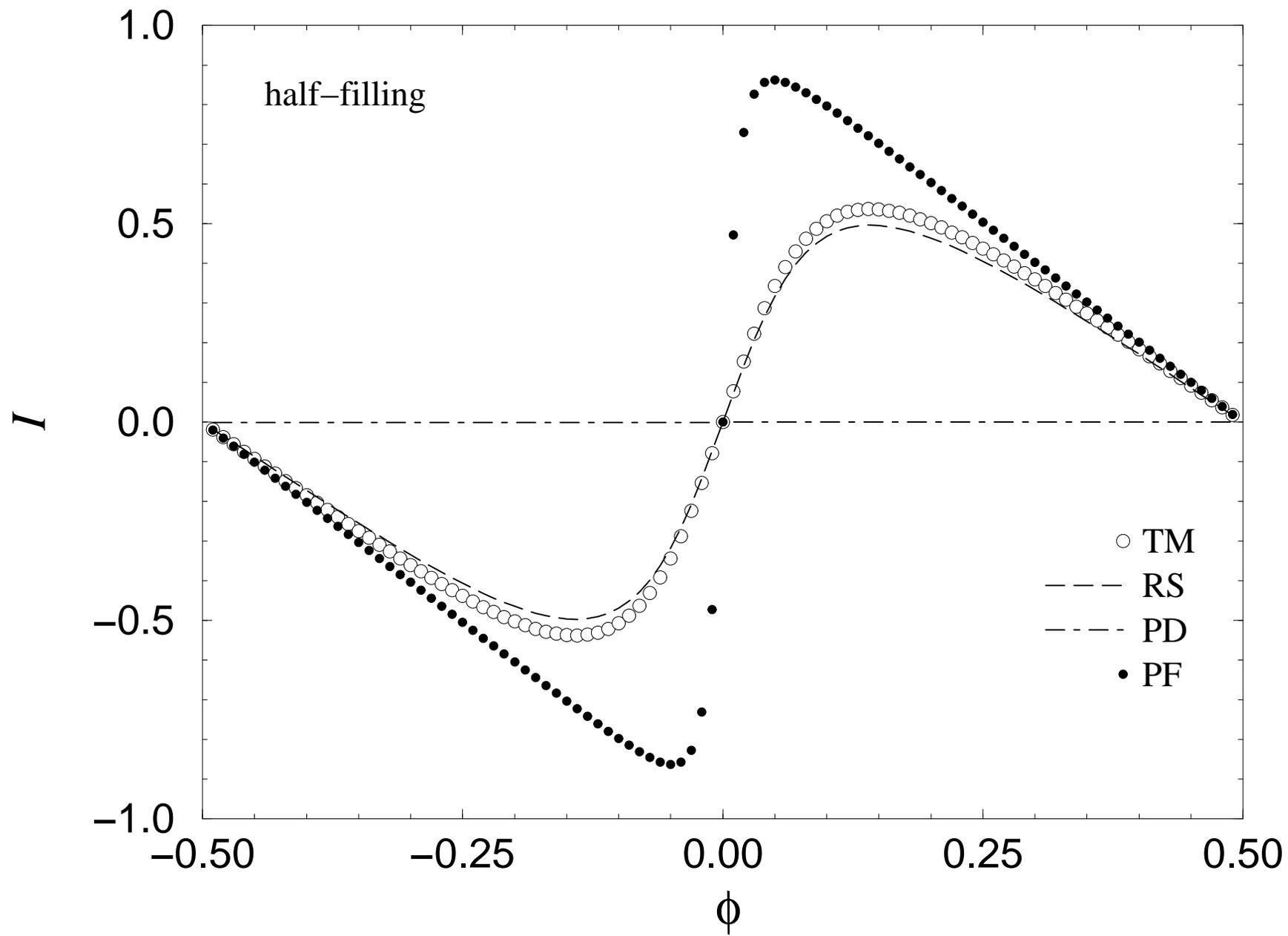
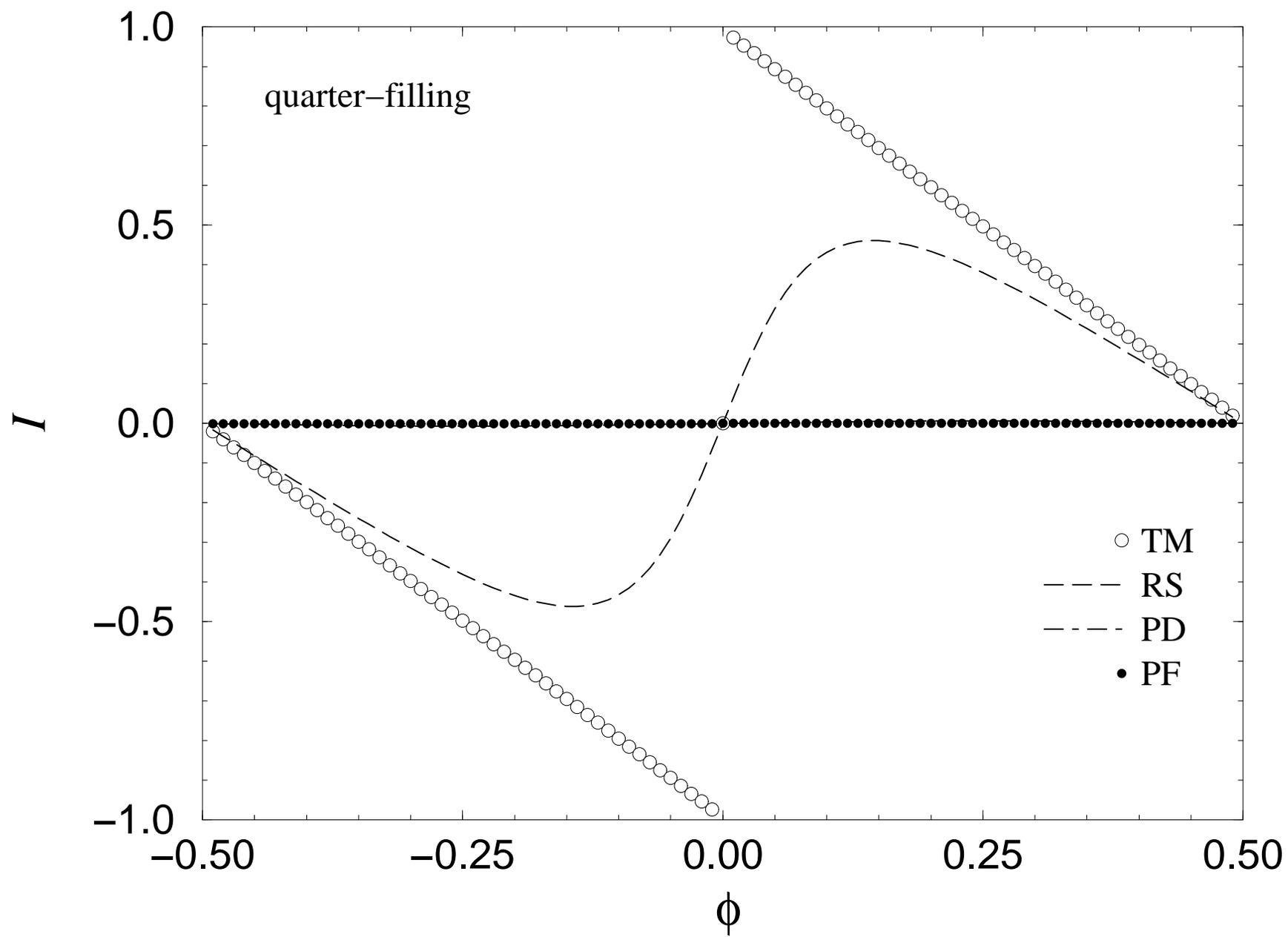
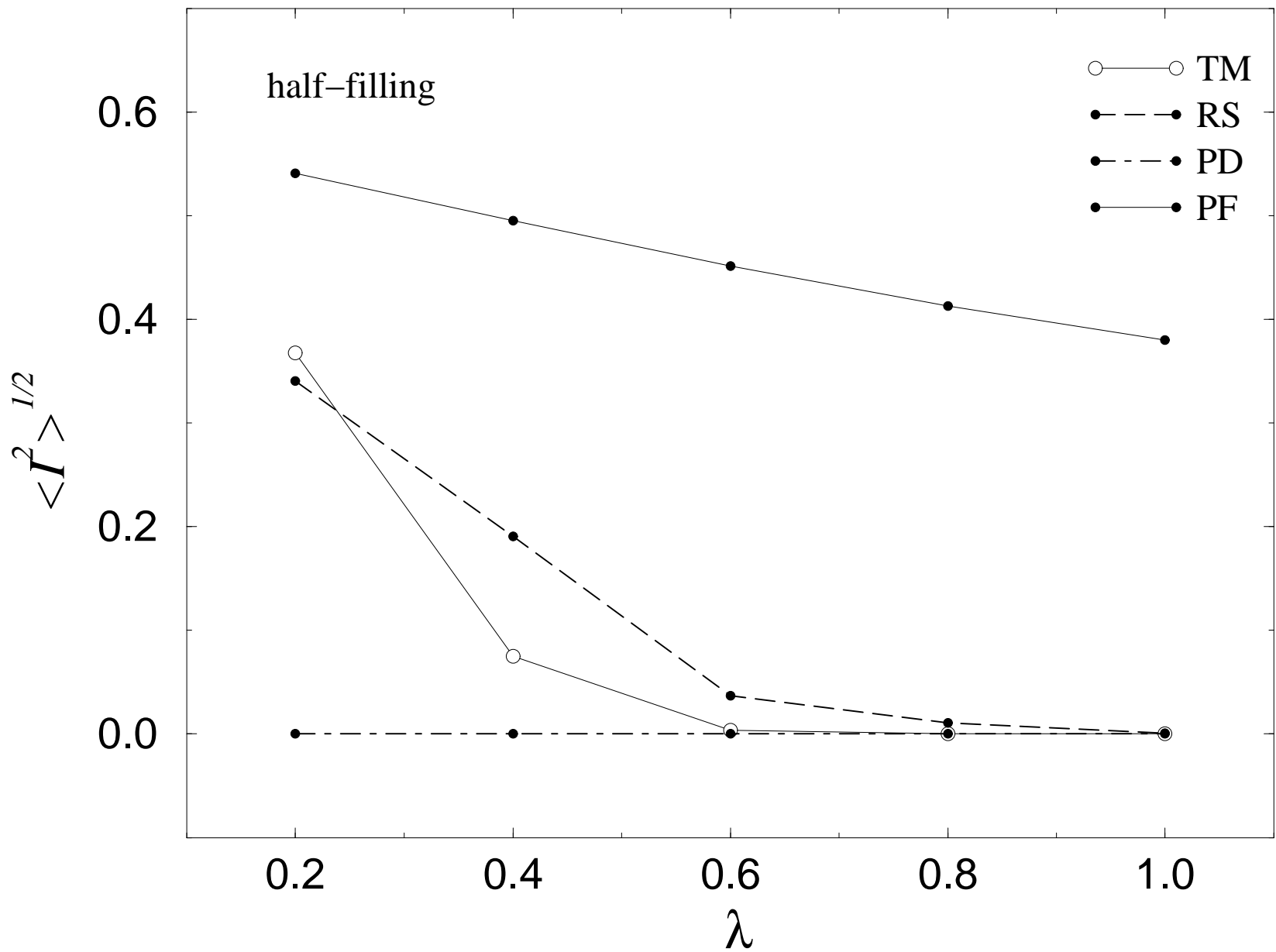
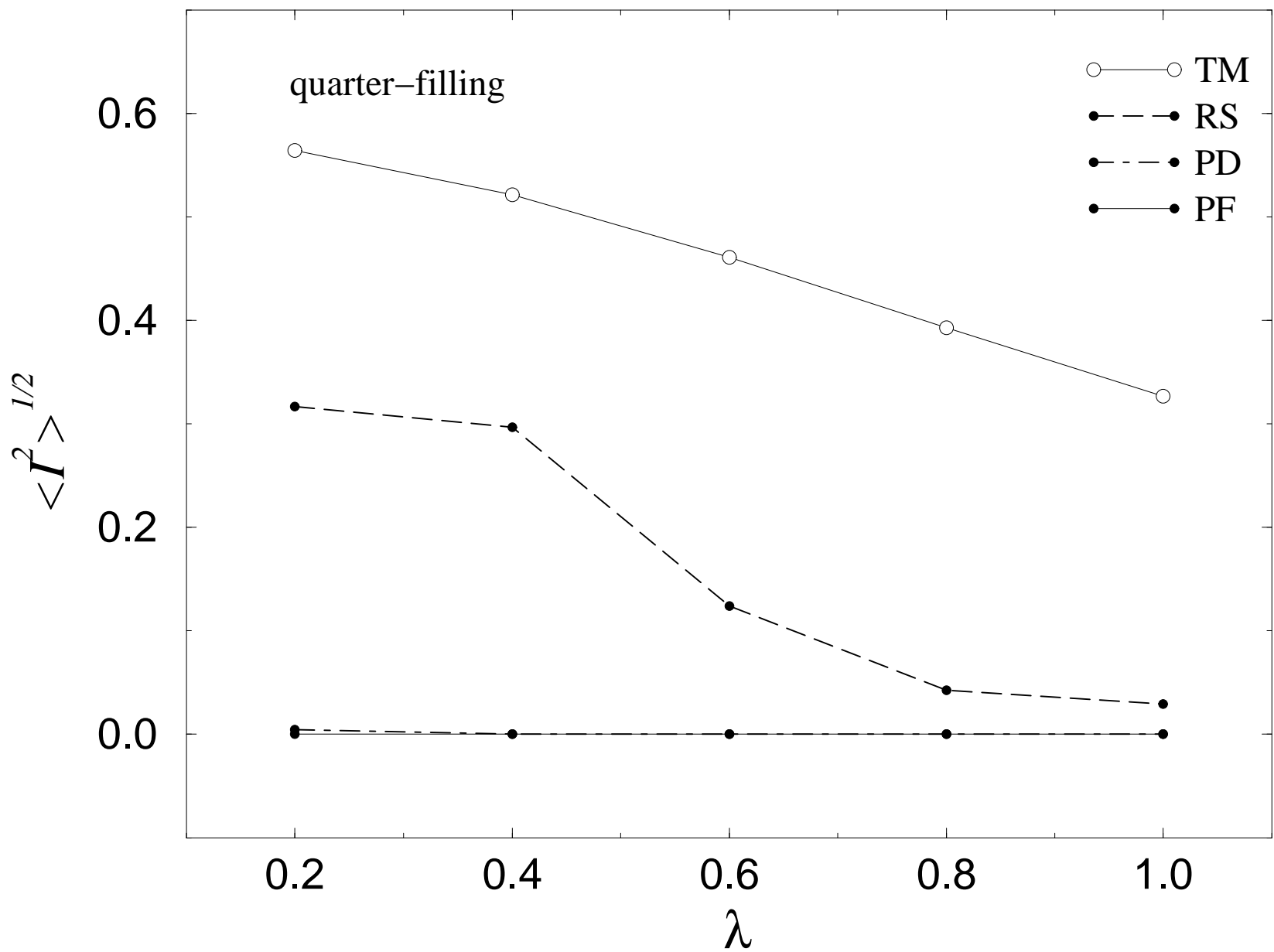
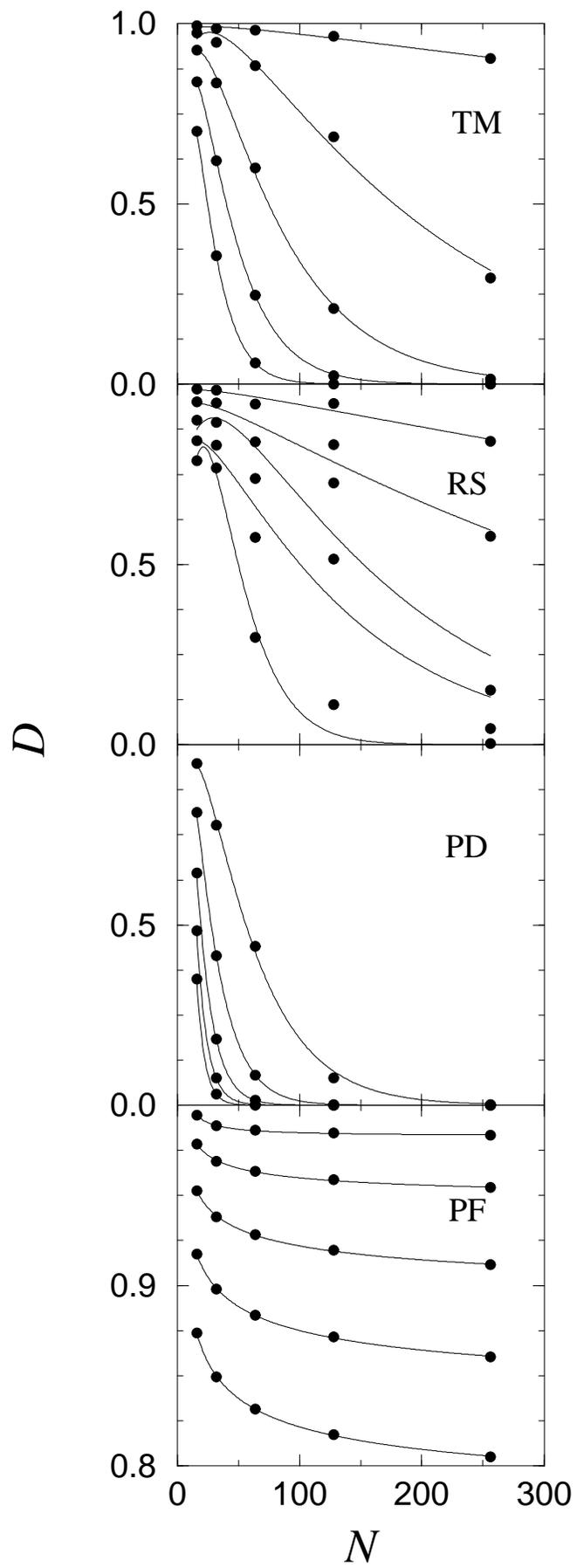
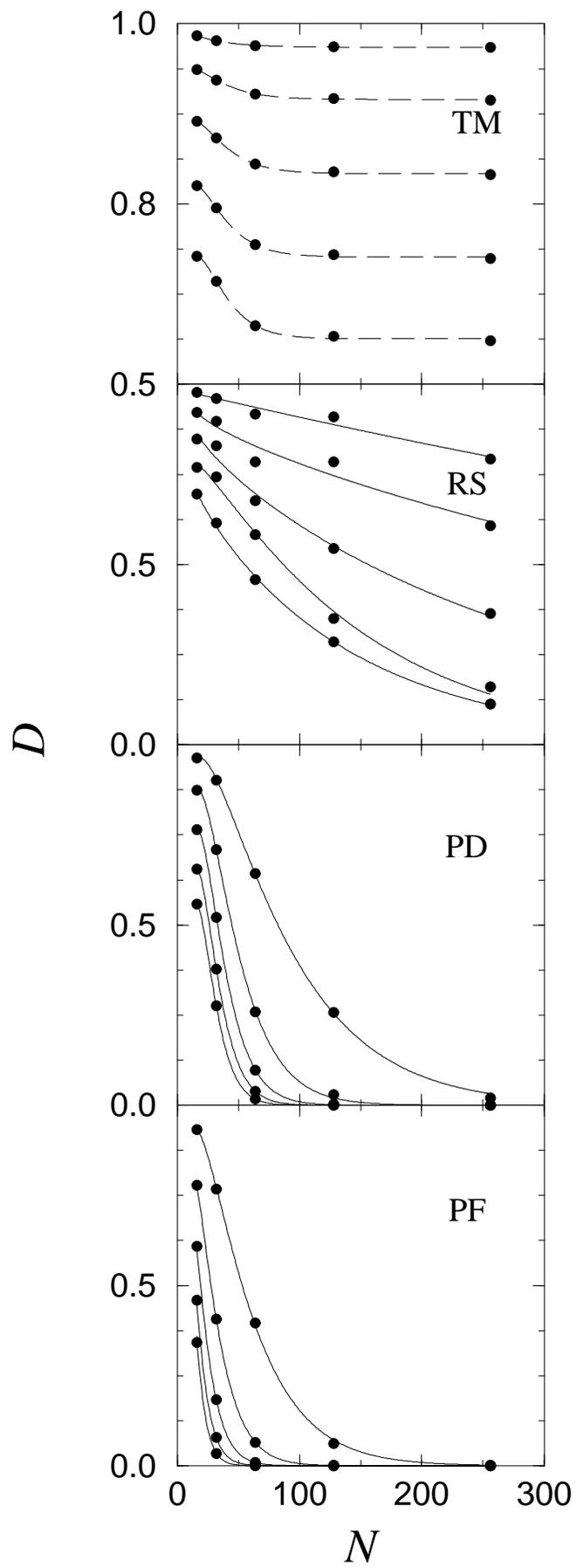
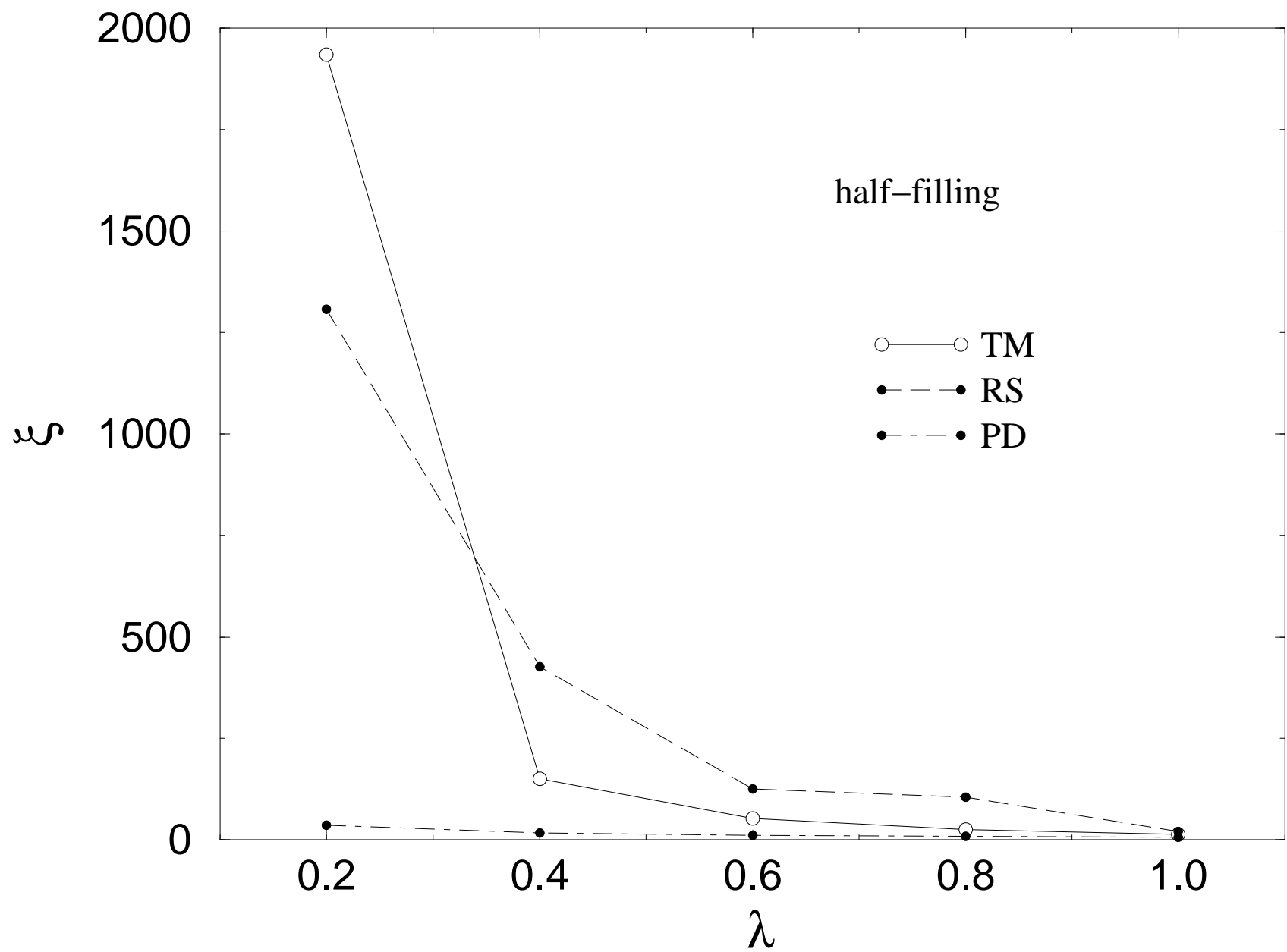
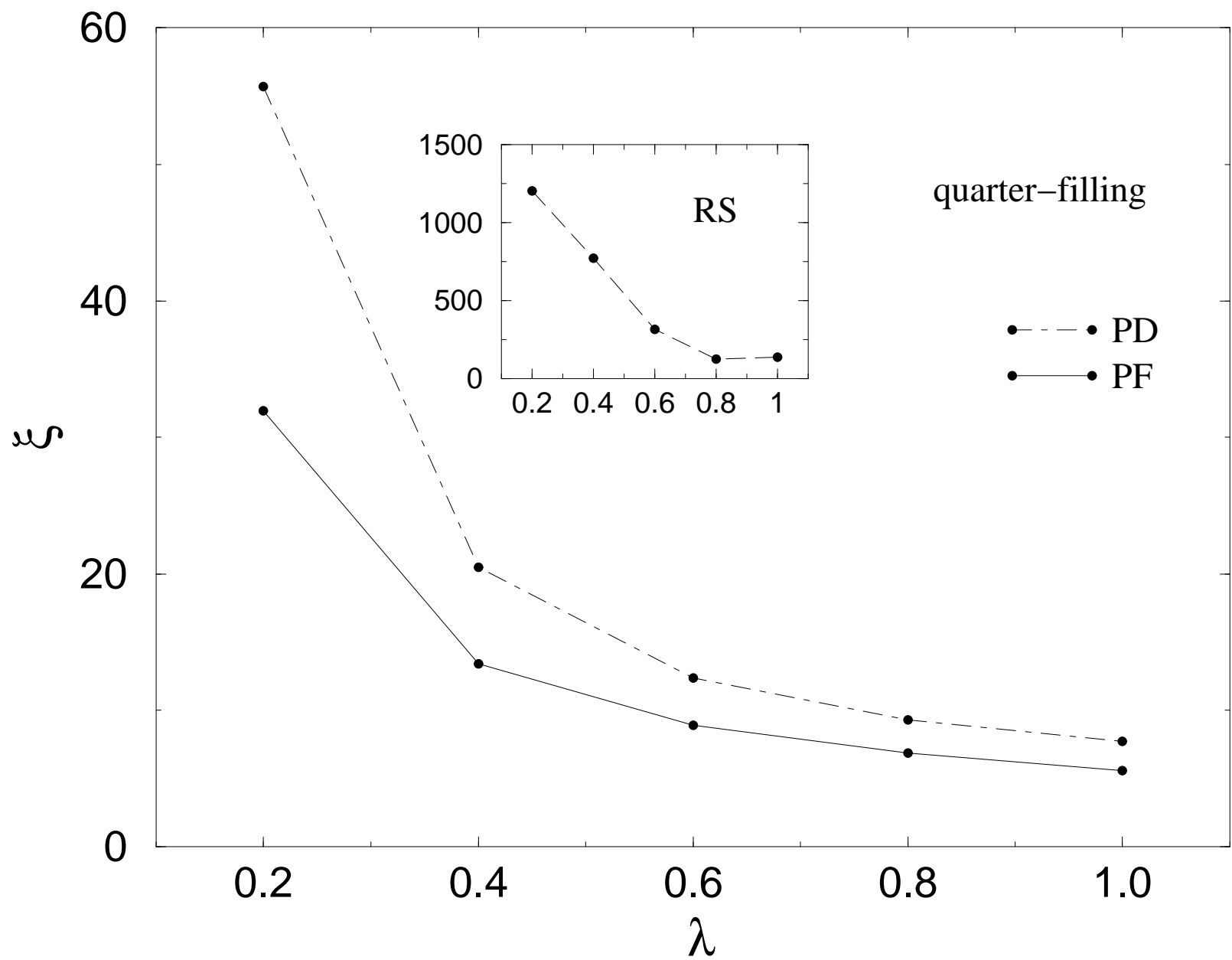
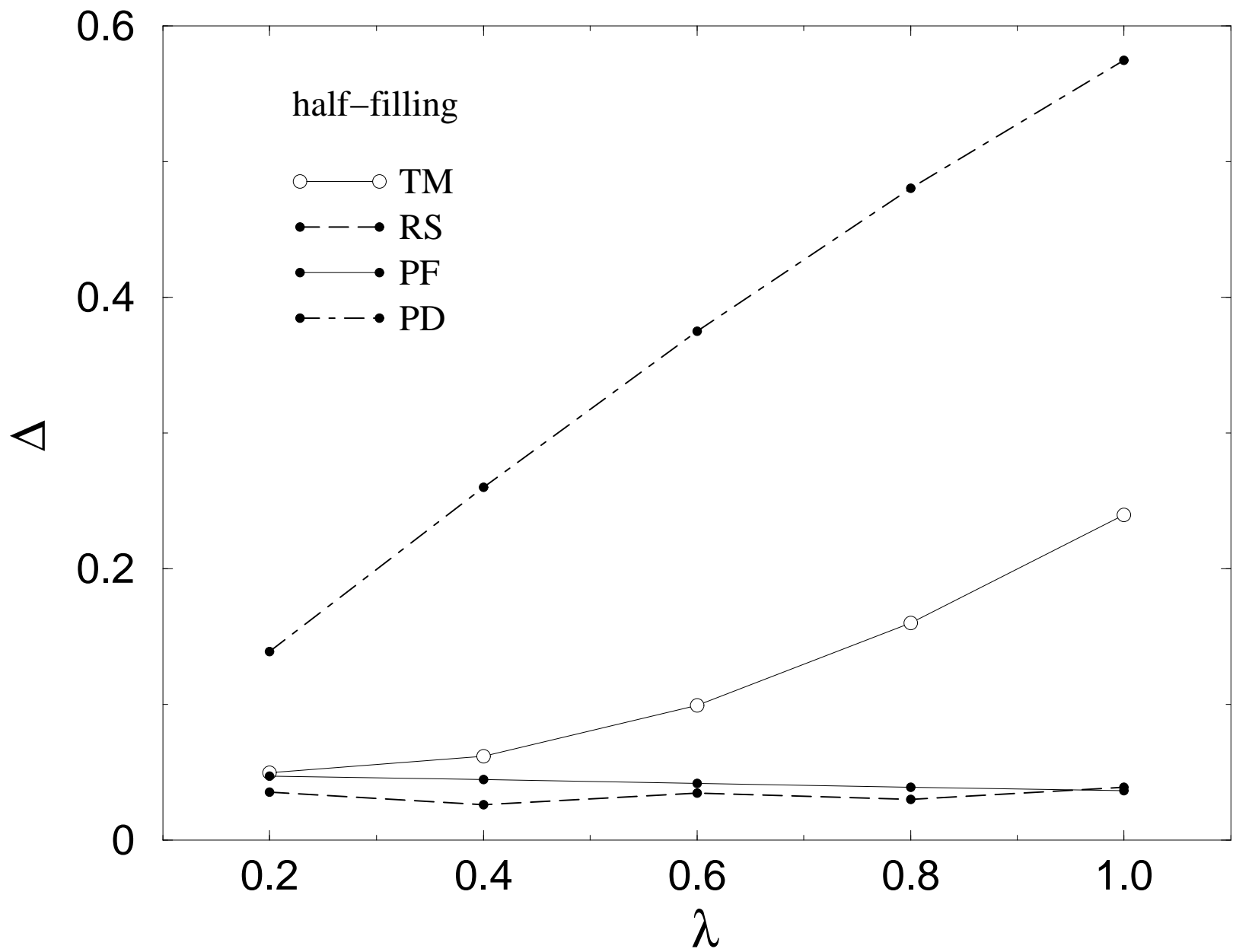
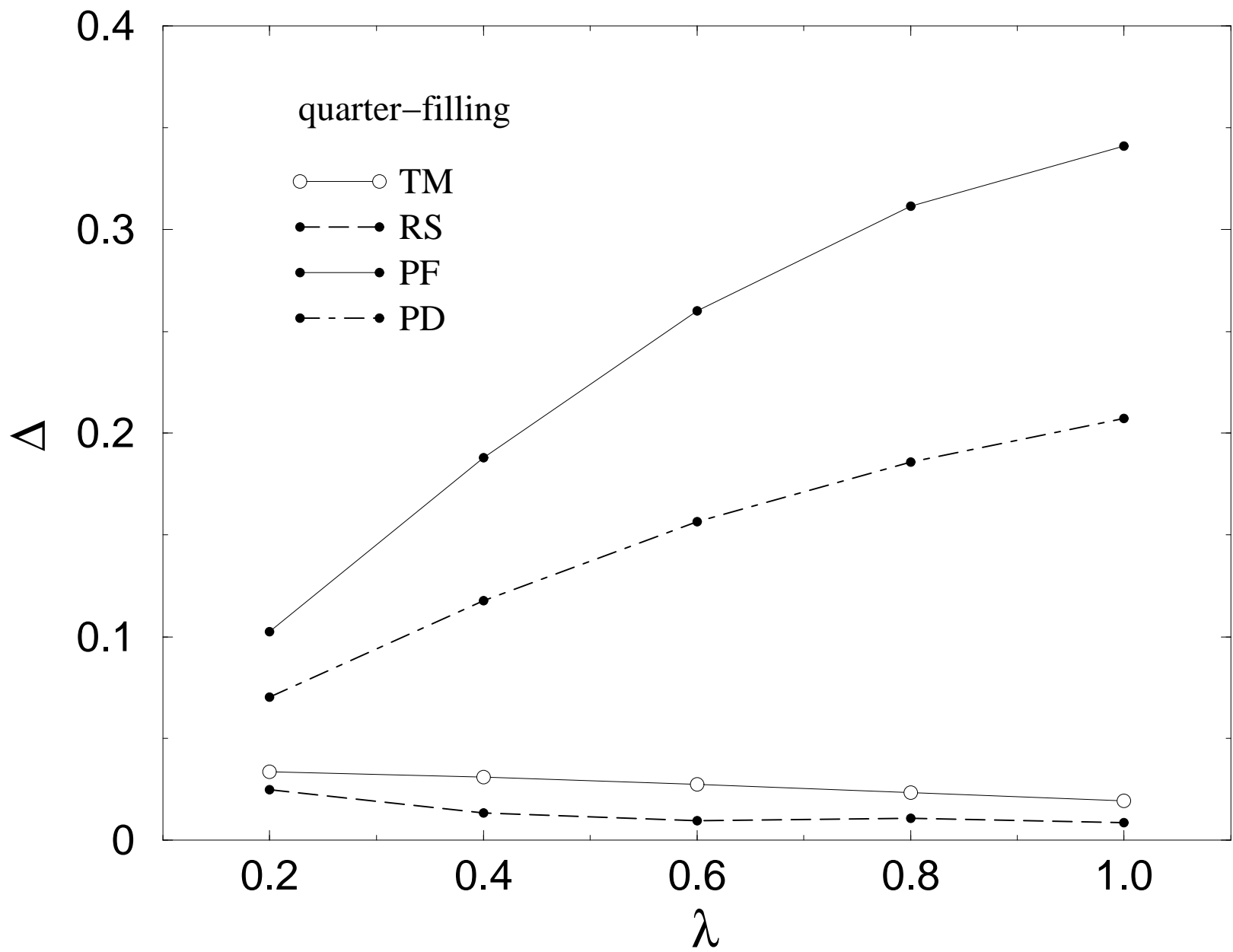
\end{document}